\documentclass[%
 reprint,
 superscriptaddress,
 amsmath,amssymb,
 aps,
 prl,
]{revtex4-2}
\pdfoutput=1

\usepackage{graphicx}
\usepackage{dcolumn}
\usepackage{bm}
\usepackage{siunitx}
\usepackage{mathtools}
\usepackage{float}
\usepackage[usenames,dvipsnames]{color}
\usepackage{wrapfig}
\usepackage{mathtools}
\usepackage{float}
\usepackage[usenames,dvipsnames]{color}
\usepackage[colorlinks=true, citecolor=blue, linkcolor=blue, urlcolor=blue]{hyperref}
\usepackage{xr-hyper}
\usepackage[printonlyused]{acronym}
\usepackage{layouts}
\usepackage{url}
\usepackage{xcolor}
\usepackage[normalem]{ulem}

\setcitestyle{super}

\newcommand{\Hamburg}{Max Planck Institute for the Structure and Dynamics of Matter, Luruper Chausse 149, 22761 Hamburg, Germany}

\newcommand{\ETH}{Institute for Theoretical Physics, ETH Zurich, 8093 Zurich, Switzerland.}

\newcommand{\PKS}{Max Planck Institute for the Physics of Complex Systems, Nöthnitzer Straße 38, 01187 Dresden,
Germany}

\begin{document}
\title{Parametrically amplified Josephson plasma waves in  $\rm{YBa_2Cu_3O_{6+x}}$: \newline evidence for local superconducting fluctuations up to the pseudogap temperature $T^*$  }

\author{Marios H. Michael}\email{mariosm@pks.mpg.de
}
\affiliation{\PKS}
\affiliation{\Hamburg}
\author{Eugene Demler}
\affiliation{\ETH}
\author{Patrick A. Lee}
\affiliation{Department of Physics, MIT, 77 Massachusetts Avenue, 02139 Cambridge, MA }

\date{\today}

\begin{abstract}
Experiments that subject underdoped $\rm{YBa_2Cu_3O_{6+x}}$ (YBCO)  to intense terahertz pulses at temperatures between the transition temperature $T_c$ and the pseudogap scale $T^*$  have revealed a reflectivity edge that resembles that of the superconducting state, together with second harmonic generation of a probe pulse  modulated at a similar frequency. These have been interpreted in terms of parametric amplification of the   lower Josephson  plasmon mode. Since this mode is often associated with coherent oscillations between bilayers in the YBCO structure, these experiments have led to the suggestion that the intense pump has created (or revealed) in-plane pair coherence up to $T^* \approx 400K$. In this paper we propose an alternative explanation by assuming the existence of local pair amplitude and phase at equilibrium for $T_c < T < T^*$. The phase correlation  spans only a few lattice constants and we do not assume  any pump-induced enhancement of this correlation, either in-plane or between  bilayers. Instead,  the coherent drive, via a parametric amplification process, induces coherence in the Josephson currents between  members of  bilayers. When combined with a Floquet framework, the reflectivity data can be explained. The key point is that in the lower Josephson plasmon, the coupling between bilayers is mainly capacitive; the Josephson current between bilayers can be set to zero without strongly affecting the parametric amplification process. Importantly, while superconducting coherence may not be created by the pump, the pseudogap phase must possess a local pairing amplitude at equilibrium. Consequently, these experiments have  strong  implications  for the understanding of the pseudogap phase.
\noindent

\end{abstract}

\maketitle
\textbf{Significance statement}
\\

\textbf{"Superconducting-like" behavior above $T_c$ has been reported when several superconducting families are subject to intense terahertz pump pulses. A well-studied system is the high $T_c$  superconductor YBCO involving pump and probe electric fields perpendicular to the plane that couple to the Josephson plasma modes. While earlier interpretations tend to suggest pump-enhanced superconductivity, we present an alternative picture that assumes the existence of highly local superconducting pairs at temperatures up to the pseudogap scale $T^*$. The coherent drives lead to coherent oscillations of the Josephson currents between members of the bilayer, without enhancing any phase coherence either in-plane or between bilayers. The notion of local pairing up to $T^*$ and Floquet-Fresnel analysis of optical properties of photoexcited states provide a unified picture of experimental data.}

\section{Introduction}
The pseudogap phase is long considered a central puzzle in the quest to understand the high $T_c$ cuprate superconductors. The pseudogap refers to the suppression of  the single particle density of states (DOS) near the Fermi level up to the pseudogap temperature $T^*$ which can be as high as 400K.\cite{keimer2015quantum} The unanswered question has to do with the origin of this suppressed DOS. Is it superconducting fluctuations that survive to several times $T_c$? \cite{emery1995importance, senthil2009coherence,lee2014amperean} Is it associated with some other order such as charge density wave or orbital current,\cite{varma2019pseudogap} or is it simply a manifestation of strong correlation physics related to antiferromagnetic spin fluctuations?\cite{wu2018pseudogap} We believe the answer may have been provided by a set of nonlinear optical pumping experiments performed on YBCO. In these experiments two phonons at frequencies 17 and 20 THz which involve out-of-plane motion of the apical oxygen are resonantly driven with terahertz short pulses with electric fields that are perpendicular to the YBCO planes. There are two kinds of observations. In an earlier set of experiments, reflectivity edges are seen in the pumped state above $T_c$ that resembles the plasmon edge at the lower Josephson plasmon frequency in the superconducting state \cite{Kaiser14}. In a second experiment, these pump pulses produce second harmonic generation (SHG) of an optical probe pulse which has been interpreted as originating from amplification of the lower branch of the Josephson plasmon \cite{vonHoegen22, michael2020parametric}.
More recently a paramagnetic signal has been reported outside a strongly driven region and interpreted as flux exclusion associated with the creation of a superconducting state by the drive\cite{Sebastian24}. In ref.~\citenum{michael2024giant} we offer an alternative explanation based on a similar set of starting assumptions used in the current paper. In this paper we will focus on how local superconducting fluctuations above $T_c$ can explain the earlier experiments on reflectivity and SHG. The relation of the present theory to the nonlinear response of YBCO under a magnetic field of Ref.~\citenum{michael2024giant,Sebastian24} is described in the discussion section.

As an aside, we note that recent studies have shown that periodic time-modulation of the superfluid density can significantly alter the electromagnetic response, even if the time-averaged density is unchanged\cite{Diessel2025,Santis2025}. The analysis in the current paper is phenomenological and we do not address the applicability of  this particular effect to the experiments on YBCO considered in this paper.
\begin{figure*}
    \centering
    \includegraphics[width= 0.98 \textwidth]{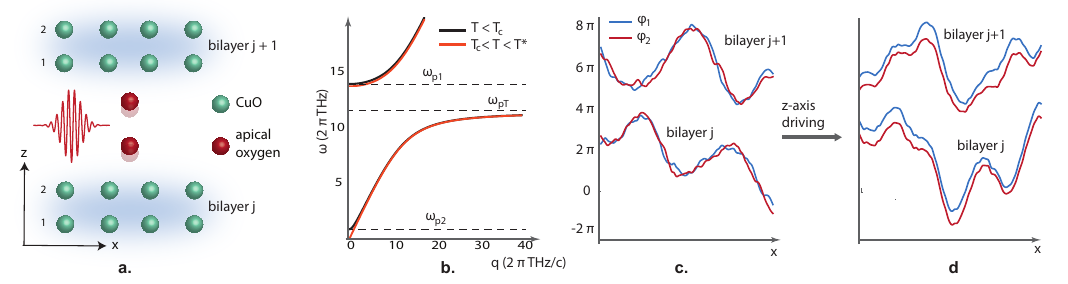}
    \caption{\textbf{Pseudogap YBCO: a.} The structure of YBCO consists of bilayers of Cu-O planes with relatively strong tunneling amplitude between them. The tunneling and therefore the Josephson coupling between bilayers is much weaker. Apical oxygen ions occupy sites between the bilayers and their vertical motion forms  two optical phonon modes that are strongly driven by the pump pulse. \textbf{b.} A sketch of the Josephson plasmon polariton dispersion showing the upper, lower and the transverse plasmon frequency respectively $\omega_{p1}$, $\omega_{p2}$ and $\omega_{pT}$. Below $T_c$ the lower plasmon polariton is gapped due to the presence of inter-bilayer coherence. Above $T_c$, in our model of the pseudogap, coherence between bilayers is lost and therefore the lower plasmon is gapless, however, the upper plasmon still exists due to short-range intra-bilayer coherence.  \textbf{c. - d.} sketches the instantaneous picture of the pair phase $\phi_{1,2}$ in bilayer $j$ and $j+1$. Red and blue lines mark the phase in layer 1 and 2 of each bilayer. \textbf{c.} At equilibrium for $T_c < T < T^*$ The \textit{relative} phase $\theta_j=\phi_{j,1}-\phi_{j,2}$ is small mod. $2\pi$ in order to take advantage of the Josephson energy while the individual phases $\phi_{j,1}$ and $\phi_{j,2}$ fluctuate wildly over many multiples of $2\pi$. \textbf{d.} Pumping produces a finite mean value of the relative phase $\theta_j$ which acquires a large time dependent value which is almost constant with respect the distance $x$ because it is synchronized to the drive. Note that $\theta_j$ is coherent even between different bilayers with different $j$ labels. This is indicated in the snapshot by placing the red curve always below the blue curve in both layers $j$ and $j+1$. On the other hand, the phases $\phi_1$ and $\phi_2$ in layer $j$ continues to fluctuate wildly and  remains uncorrelated with the phases $\phi_1$ and $\phi_2$ in layer $j+1$. }\label{fig:plasmon}
\end{figure*}

Recall that the YBCO structure consists of  bilayers formed out of two closely spaced copper-oxygen planes, but the bilayers are quite far separated from each other (see Fig.~\ref{fig:plasmon}a.) At equilibrium in the superconducting state, the higher frequency Josephson plasmon at 14.2 THz involves mainly intra-bilayer charge oscillations while the low frequency one at 0.9 THz involves mainly inter-bilayer oscillations (see dispersion in Fig.~\ref{fig:plasmon}b.). In earlier papers ~\cite{vonHoegen22, michael2020parametric} the interpretation of the pumped state was that the  driven phonon couples to a pair of plasmons, one at the upper and one at the lower branch, conserving total energy and momentum. 
This process can be understood as time periodic modulation of the index of refraction \cite{michael2020parametric}. More recently a two dimensional pump-probe experiment indicates that the drive field excites two phonon modes at ~17 at 20 THz and it is the difference frequency that excites a pair of lower plasmons. \cite{taherian24}    Theories based on either of these  models can explain  the experimental observation that  the photoinduced edge in the reflectivity is higher than the low temperature Josephson plasmon edge. The more recent model matches this frequency to half of the frequency difference of the IR phonons excited by the pump pulse \cite{taherian24}.

The SHG experiment is even more direct because the amplified Josephson plasmon mode is directly observed.  While this picture successfully accounts for much of the observations, it gives the impression that since the inter-bilayer plasmons are strongly excited, it is necessary that phase coherence between the bilayers on the lengtscales comparable to the wavelength of excited plasmons, i.e. tens of microns, was present before the pump pulse or has been induced by the pump pulse. This in turn led to the suggestion that the drive has revealed (or created) coherence within each Cu-O plane on the same lengthscales. 
 \cite{vonHoegen22, taherian24,michael2020parametric}. This explanation leaves several open questions as to how this enhanced pairing state can come about due to driving, or whether such coherence is necessary to explain the data in the first place.

The goal of this paper is
to formulate a model that explains the data while making the \textit{minimal} assumption for the coherence of the pairing order parameter
in the equilibrium state before the drive, and still
retain a theory that explains the data. Furthermore, we assume that superconducting coherence is not modified by the pump pulse. 
Our model is as follows:

1. We need to assume that at equilibrium, a local pairing amplitude exists so that a phase of the local pairing order parameter, $\Delta_{\alpha,j}(\vec{r}) = | \, \Delta_{\alpha,j} (\vec{r})\, | \, e^{ i\phi_{\alpha,j}(\vec{r})}$, is defined. Here $\alpha = 1,2$ labels the top and bottom layer within a  bilayer labeled by $j$, and $\vec{r}$ is the in-plane coordinate. The phase $\phi_{\alpha,j}$ has short range order with correlation length $\xi$ which can be very short, as little as a few times the zero temperature coherence length $\xi_0$ of the superconductor which is about 2 nm. We require the correlation length  $\xi$ to be larger than $\xi_0$ for the local pairing amplitude to be established, but it cannot be too large as temperature increases from $T_c$ towards $T^*$ in order for observables such as fluctuation conductivity and diamagnetism to be small enough to be unobservable. On the other hand, the correlation between members of a bilayer can be much longer. This is because  the Josephson coupling between layers 1 and 2 within a given bilayer in YBCO is quite strong, enough to give  significant correlation between $\phi_{1,j}$ and $\phi_{2,j}$ up to relatively high $T$. This is shown schematically in Fig. \ref{fig:plasmon}c. In  a more microscopic cartoon picture, we consider an intermediate temperature range where the coherence length is long enough such that free vortices are identifiable in each layer as in the standard Berezinskii, Kosterlitz, Thouless (BKT) formalism, but the vortices are bound between layers 1 and 2. These vortices lead to short range order in the phases $\phi_1$ and $\phi_2$ of each layer, but does not disorder the relative phase $\phi_1-\phi_2$. This kind of intra-bilayer coherence will give rise to a transverse Josephson plasmon mode.  There is evidence for the existence of this mode in optical conductivity up to 180K\cite{dubroka2011evidence}. However, up to now there has been no direct evidence for pairing amplitude that survives up to $ T^* \approx 400K$. So our assumption is untested, but at the same time, if verified, has an important impact on our understanding of the pseudogap phase.


2. The interlayer Josephson coupling energy scales with the square of the Josephson plasmon frequency. In YBCO at low temperatures, the frequency ratio between the lower and upper Josephson plasmons is approximately an order of magnitude, implying that intra- and inter-bilayer couplings differ by a factor of 100. Consequently, we assume that Cooper pair tunneling between bilayers is negligible within the pseudogap regime.

3. In view of ref.\citenum{taherian24}, we assume that the  drive couples to two phonons at $\approx$17 and 20 THz and the difference frequency  parametrically couples to two lower plasmon modes that are derived under condition 1 above. This is illustrated in Fig. \ref{fig:Parametric}a,b. 

We emphasize that no inter-bilayer coherence is needed for the existence of these modes. To prove this point, we can set the Josephson coupling between bilayers to zero and still recover a low frequency mode which is very similar to the inter-bilayer plasmon mode that was derived assuming inter-bilayer phase coherence. This is because the bilayers are coupled capacitively, so that the mode does not require physical currents between bilayers. As seen in Fig.\ref{fig:plasmon}b, the main difference between the two cases is that with long range coherence the lower plasmon mode has a small gap (the lower plasmon frequency of $0.9$ THz), whereas for decoupled layers we have a linearly dispersing mode as $q \rightarrow 0$. At finite $q$ and $\omega$ the difference between the two dispersions is minimal. We suggest that the external drive creates a coherent excitation in the relative phase between bilayers, where the capacitive coupling is responsible for forcing $\phi_{1,j}-\phi_{2,j}$ to be aligned with $\phi_{1,k}-\phi_{2,k}$ even when $j$ and $k$ are far apart.
However, there is no enhanced coherence in the phase $\phi_{1,j}$ itself, whether within a layer or between bilayers. This is illustrated schematically in Fig.\ref{fig:plasmon}d.

Our main conclusion is that with the new level of understanding, we believe the Josephson plasmon parametric amplification picture is the correct explanation of the experiment reflectivity and the SHG experiments. However, the implication is that pairing correlation with very short range order persists up to $T^*$ at equilibrium, hence the pseudogap co-exists with or is tied to fluctuating superconductivity.

Below we give the details that lead to our conclusions. First we derive the collective modes under condition (1) based on a new formulation using electric and magnetic field directly instead of gauge potential as was done in Ref.~\citenum{michael2020parametric}. The treatment is classical and is simpler, so that the interpretation is more transparent. An important physical point is that the drive couples to the intra-bilayer Josephson current which depends on a phase difference within a bilayer, and not the phase itself.  Then we discuss the coupling of the Josephson plasmons to the phonon drive and their parametric amplification. This part is closely related, albeit with some modifications, to what was done earlier in Ref.~\citenum{michael2020parametric}. 
Our analysis does not provide any information on the nature of the pairing, whether it is d-wave or pair density wave, for instance. We also do not make a statement concerning the origin of the energy gap associated with the pseudogap itself, which is often take to be of order  100 meV. Our theory is phenomenological and does not depend on an understanding of the microscopic picture. Nevertheless, the notion that pairing amplitude persists up to 400K in YBCO is a significant and somewhat provocative claim. These and related questions will be further discussed in the Conclusion section.






\section{The Josephson plasmon polariton in bi-layer YBCO}
The Josephson plasmons are admixed via the Maxwell equation to the propagation of light, and is more properly called the Josephson plasmon polariton (JPP) modes.
In this section we give results of the JPP modes for a structure consisting of stacked bilayers that is appropriate for YBCO. We determine the mode structure by solving the Maxwell equations with the assumption that the in-plane currents and charge densities are confined within the cuprate layers while Josephson currents flow between the layers. It is worth noting the difference from the earlier treatment given by Michael et al~\cite{michael2020parametric}. There the Maxwell equation was formulated in terms of the vector potential $\textbf{A}$ and the approximation was made that the in plane component of $\textbf{A}$ is confined to the layers. Here we do not make this assumption. It turns out that the formulation in terms of $\textbf{E}$ is more amenable to analytic treatment and we can solve the full three dimensional structure of $\textbf{E}$ for the case when the momentum perpendicular to the plane $q_z$ is zero. Details are shown in Appendix A and B. We demonstrate that for YBCO, the detailed structures of the x component of the electric field treated in Appendix B can be ignored, and we recover the standard expressions for  the upper and lower JPP modes $\omega_U$ and $\omega_L$ 
\begin{widetext}
\begin{align}
\label{dispersion1}
    \omega_{U,L}^2= (\omega_{p1}^2+\omega_{p2}^2+c^2q^2)/2 \pm 1/2 \sqrt{c^4q^4+2c^2q^2(\omega_{p1}^2+\omega_{p2}^2-2\omega_{pT}^2)+(\omega_{p1}^2-\omega_{p2}^2)^2} 
\end{align}
\end{widetext}
The Josephson plasmon frequency $\omega_{p1}$ for slab $1$ (the region of space between layers that form the  closely spaced bilayers) is given by
\begin{align}
    \omega_{p1}^2= e^* j_{c,1} d_1.
\end{align}
where $e^*=2e $, $ j_{c,1}$ is the amplitude of the Josephson current in slab 1   and similarly for slab 2 (the region between bilayers). The thickness of Slab 1, (2) is   given by $d_1$,  ($d_2$) and $D=d_1+d_2$.
The transverse plasmon frequency is given by
\begin{align}
\label{transverse}
   \omega_{pT}^2=(d_1/D)\omega_{p2}^2+(d_2/D)\omega_{p1}^2
\end{align}

A sketch of the JPP dispersion is shown in fig. \ref{fig:plasmon}b. Here we make a number of observations that will have important consequences for the main point of this paper:

1. As long as $\omega_{p2} \ll \omega_{p1} $, its effect on the dispersion is minimal and limited to a region in momentum space where $cq \approx \omega_{p2}$, as seen in Fig. \ref{fig:Parametric}b. To emphasize this point, we can simply set $\omega_{p2}=0$. This will decouple the phase correlation between bi-layers, leading to a phase distribution sketched in Fig. \ref{fig:plasmon}c. The reason the phase decoupling has little  effect on the plasmon polariton spectrum is that the coupling between layers is mainly a capacitive coupling. In the solution of the Maxwell equation, no physical current is required between bi-layers. The source of the electromagnetic field is  Maxwell's displacement current.

2. If we set $\omega_{p2}=0$, the lower plasmon is gapless with a linear dispersion. However its velocity is given by 
$c'=\sqrt{\frac{d_2-d_1}{d_2+d_1}}c$, i.e. it is different from propagation in free space and this modification of  the velocity exists only because $\omega_{p1}$ is nonzero. Therefore even the linear dispersion knows about the phase difference between members of a bilayer and this linear mode is coupled to the Josephson current between them.

3. We can solve for the ratio between the $z$ components of the electric field $E_1$ and $E_2$ in slab 1 and 2 at a given mode frequency $\omega$. (See appendix A) 
\begin{align}
\label{transverse}
   \frac{E_1}{E_2}=\frac{\omega_{p2}^2-\omega^2}{\omega_{p1}^2-\omega^2}
\end{align}
It is easy to see  that the $z$ component of the electric field is nonzero in both slabs, but is mainly in slab 1 for $\omega_U$ and in slab 2 for $\omega_L$. Since the mode is a linear superposition of $E_1$ and $E_2$, we can use this ratio to estimate the coupling of this mode to a drive that couples to the intra-bilayer electric field $E_1$.  Note that while this coupling is small, it is the same order of magnitude whether $\omega_{p2}$ is zero or not for the frequencies we are considering.

Finally we write down the well-known expression for the dielectric function $\epsilon (\omega)$:
\begin{align}   \label{epsilon}
    \epsilon (\omega)= \frac{(\omega^2-\omega_{p1}^2)(\omega^2-\omega_{p2}^2)}{\omega^2 (\omega^2-\omega_{pT}^2)}
\end{align}
From this expression, the plasmon dispersion given by Eq.\ref{dispersion1} can be obtained by the solution of the equation $q^2 = \epsilon(\omega) \omega^2$. The upper and lower plasmons $\omega_{U,L}$ are the zero's of $\epsilon(\omega)$ while the transverse plasmon $\omega_{pT}$ is the pole. The former gives rise to the plasma edges in the reflectivity while the latter gives a peak at $\omega_{pT}$ in the absorption, i.e. in $\sigma_1(\omega)$. \cite{dubroka2011evidence} The latter is often taken to be a signature of superconductivity. It is worth noting that the transverse plasmon peak has been seen in underdoped YBCO up to 180K. \cite{dubroka2011evidence} This is direct evidence that some pairing exists up to that temperature scale. However, it is important to remember from the discussion above that no inter-bilayer coherence is required for the formation this mode. Therefore the existence of this mode provides information only on the coherence between members of the bilayers. Furthermore, this coherence does not need to be long range: a finite correlation length will just give rise to a broadening of this mode, leading to a loss in intensity.

\section{Parametric driving of the lower Josephson polariton}
\label{ParametricDrivingSection}

Now we describe our model for the pumping process and the resulting instability due to parametric amplification of the lower Josephson plasmon polariton mode. We assume that the THz drive pumps two optical phonons, leading to oscillations at their fundamental frequencies $\omega_{ph,1} = 17 \ THz$ and $\omega_{ph,2}=20 \ THz$ with some damping : $Q_1 = \theta(t)e^{-\gamma t} \sin (\omega_{ph,1} t) $ and $Q_2 = \theta(t)e^{-\gamma t} \sin (\omega_{ph,2} t) $. The phonon oscillation modulates  the Josephson coupling between members of a bilayer, leading to a time dependent nonlinearity in the Josephson coupling: 
\begin{eqnarray}
j_{c,1}(t) = j_{c,1} \left(1 + \alpha \left( c_1 Q_1 + c_2 Q_2 \right)^2 \right).
\label{JcQ2}
\end{eqnarray}
This model (also called four-wave mixing) was introduced in ref.~\citenum{michael2020parametric}, but the treatment there focused on an alternate three-wave mixing model. A more recent experiment supports the four-wave mixing scenario \cite{taherian24, liu2024probing, gomez2024principles} and this will be our focus in this paper.

\begin{figure*}
\centering
\includegraphics[width=0.99\textwidth]{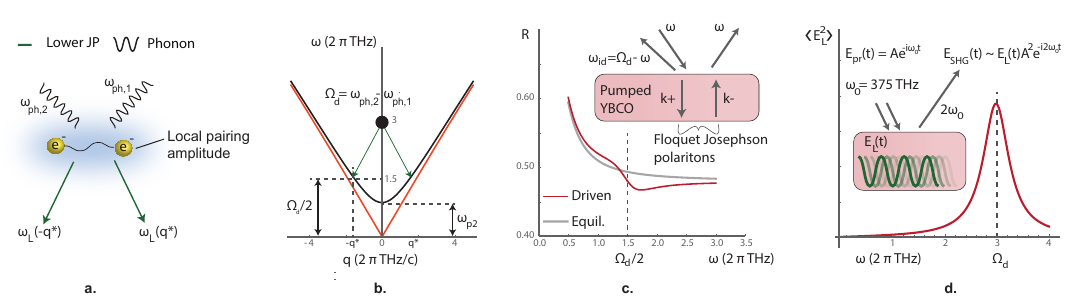}
\caption{\textbf{Parametric driving of Josepshon plasma waves: a.} Schematic of the four wave mixing process between the two apical oxygen phonons and lower Josephson plasmons. We show that local pairing amplitude enables a modulation of the Josephson coupling between members of a bi-layer and 
is sufficient to generate pairs of lower Josepshon plasmons, even when there is no superconducting coherence between different bi-layers. \textbf{b.} An expanded view near the lower Josephson plasmon polariton. Red line indicates the case when inter-bilayer Josephson coupling has been set to zero. The drive frequency $\Omega_d$ which is equal to the difference of the two phonon frequencies leads to the parametric generation of two lower plasmon modes with opposite momenta, which corresponds to creating a plasmon squeezed state with $\langle a(q) a(-q) \rangle \neq 0$. The amplification of the resonant modes and their momenta, $\pm q^*$, do not depend much on whether $\omega_{p2}$ is finite or zero.
\textbf{c.} We plot the equilibrium and driven reflectivity that arises in the pseudogap YBCO. The equilibrium reflectivity is a sketch meant to mimic experimental data in Ref.~\cite{vonHoegen22}. The driven reflectivity is computed using the universal formula for Floquet reflectivity found in Ref.~\citenum{Marios22}, $R_{driven} = R_{eq} + C Re[\frac{-e^{i \theta}}{w - \Omega_d/2 + I \gamma}]$, with parameters $\Omega_d = 3$ THz, $C = 0.01$, $\theta =0.4$ and $\gamma = 0.3$ THz. The inset shows a sketch of the reflectivity process from a driven medium: Light in the driven medium propagates through two Floquet Josephson polaritons oscillating at both $\omega$ and $\omega - \Omega_d$ with momenta $k_+$ and $k_-$. Light is reflected at both $\omega$ and $\Omega_d - \omega$, which is computed by matching Floquet-Fresnel boundary conditions. \textbf{d.} Plot of the Second Harmonic Generation intensity as a function of time. The amplified lower Josephson plasmon breaks inversion symmetry and produces SHG of the probe pulse at $\omega_0=375 \ THz$. The SHG amplitude is proportional to coherent oscillations of the lower Josephson polariton, $E_L(t)$ while its intensity, $I_{SHG}$ is proportional to $ \langle E_L^2\rangle(\omega)$. In the inset it shows that coherent lower Josephson polaritons excited due to the drive are seeded by thermal fluctuations and will have random phases of $a(q)$ and $a(-q)$ individually. However, their product $\langle a(q) a(-q) \rangle  $ has a phase fixed by the parametric drive. The intensity of SHG is only sensitive to the combined product $\langle a(q) a(-q) \rangle  $ which has a fixed phase. Thus, it oscillates at $\Omega_d = 3 $ THz. The plot is sketching Lorentzian profile centered at $\Omega_d$ with a width of $0.6$ THz.}
\label{fig:Parametric}
\end{figure*}

The pump modulates 
the Josephson plasmon frequency $\omega_{p1}$:
\begin{align}
    \omega_{1,dr}^2(t) =& \omega^2_{p1}\left( 1 + \lambda_{JP,ph} Q_1 Q_2 \right), \\
\end{align}
which parametrically amplifies  a JPP mode at momentum $q^*$ when $\omega_{ph,2} - \omega_{ph,1} = 2 \omega( q^*)$. The four wave mixing process is demonstrated schematically in Fig.~\ref{fig:Parametric} a. and b. Amplification of this mode  appeares experimentally as features in the reflectivity at $\omega \sim 1.5$ THz and in SHG at 3 THz, as shown in Fig.~\ref{fig:Parametric} c and d, and explained below. 
To give a bit more detail, in terms of the electric field along the z-direction, $E_1$ and $E_2$, the equations of motion (setting $\omega_{p2}=0)$ are: 

\begin{equation}  
  \frac{1}{c^2}\partial_t^2 E_{1} - \frac{d_1 }  {d_2 + d_1 }\partial^2_x E_{1} + \frac{\omega^2_{p1}}{c^2} E_{1} - \frac{d_2}{d_2 + d_1 } \partial_x^2 E_{2} = 0 , 
  \label{eq25}
\end{equation}

\begin{equation}
  \frac{1}{c^2}\partial_t^2 E_{2} - \frac{d_2 }{d_2 + d_1 } \partial^2_x E_{2} - \frac{d_1}{d_2 + d_1 } \partial_x^2 E_{1} = 0.
\label{eq26}
\end{equation}
The coupling energy takes the form $\sim E_1^2(x)Q_1Q_2$.
By going to Fourier space, we find that $E_{1,2}(q,\omega)$ is coupled to $E_{1,2}(-q,\omega+n\Omega_d)$. Withing degenerate Floquet perturbation theory\cite{Eckardt_2015, Marios22,nazaryan2024} we can truncate the 
Floquet equations of motion to n=-1. We find: 
\begin{widetext}
\begin{equation}
   \begin{pmatrix}  -\omega^2 + \omega_{p1}^2 + v^2_{1}q^2 &&v_2^2 q^2 && \alpha \omega_{p1}^2 && 0  \\
   v_1^2 q^2 &&  - \omega^2 + v^2_2 q^2  && 0 &&0 \\ 
   \alpha \omega_{p1}^2 && 0 && -(\omega - \Omega_d)^2 + \omega_{p1}^2 + v^2_{1} q^2 && v^2_{2}q^2 \\
   0 && 0 && v^2_{1} q^2 && -(\omega - \Omega_d)^2  + v^2_{2} q^2   \end{pmatrix} \cdot \begin{pmatrix} E_1(q,\omega) \\ E_2(q,\omega)\\ E_1(-q,\omega- \Omega_d) \\ E_2(-q,\omega- \Omega_d)\end{pmatrix}  = 0 ,
\end{equation}
\end{widetext}
where $\Omega_d =\omega_{ph,2} - \omega_{ph,1} = 3$ THz, and $v_1^2=(d_1/D)c^2$, $v_2^2=(d_2/D)c^2$. 
The solution of this equation leads to unstable modes with imaginary frequency.
We can simplify the problem by focusing on a pair of eigenstates in the lower JPP branch, $E_{L,q}$, with frequency $\omega_{L,q}$ at $\pm q$ with energy close to $\Omega_d/2$ and project to a 2x2 matrix
\begin{equation}  \label{matricx2by2}
   \begin{pmatrix}  -\omega^2+\omega^2_{L,q} &&\lambda  \\
   \lambda^* && -\omega_{id}^2 + \omega^2_{L,q}   \end{pmatrix} \cdot \begin{pmatrix} E_{L,q}(\omega) \\ E_{L,q},-\omega_{id})\end{pmatrix}  = 0 ,
\end{equation}
where we defined the idler frequency $\omega_{id} = \Omega_d - \omega$. Eq.~\ref{matricx2by2} takes the form of the standard matrix for parametric amplification. The solution for $\omega$ near $\omega_L(q)$ is given by
\begin{align}   \label{epsilon}
    \omega=\Omega_d/2 \pm \frac{1}{2}\sqrt{\delta^2-\frac{|\lambda|^2}{2\omega_L(q)(\Omega_d-\omega_L(q))}}
\end{align}
where $\delta=\Omega_d-2 \omega_L(q)$ is the detuning parameter and $\lambda$ is the effective coupling which is related to $(E_1/E_2)$  as discussed earlier. This process will pick out a resonant mode when $\Omega_d=2 \omega_L(q)$ and amplified it with exponential growth, as illustrated in Fig. \ref{fig:Parametric}b.

\section{Implications of parametric driving of Josephson polaritons for experiments}
\subsection{Time-resolved Second Harmonic generation}
The parametric amplification explains the second harmonic generation observed in the experiment. The oscillation of the amplified $\omega_L$ mode breaks the inversion symmetry which is needed to produce an SHG signal. The SHG susceptibility that relates the SHG signal, $E_{SHG}$, at  $2 \omega_{0}$ to the probe electric field, $E_{pr}(t) = A e^{- i \omega_0 t}$, at the probe frequency $\omega_{0}$ is therefore proportional to the electric field associated with the amplified lower plasmon mode $E_L$ which oscillates at $\Omega_d/2$: 
\begin{equation}
    E_{SHG} \propto E_L(t)\times  E_{pr}^2 e^{2 i \omega_0 t},
\end{equation}

While we framed this process as a parametric instability process at $\omega_L(q^*) = \Omega_d/2$, even in the presence of sufficient dissipation to prevent this instability, the fluctuations of $E_L(t)$ are strongly modified due to the parametric resonance. In fact in ref.~\citenum{taherian24}, it is shown that $\langle E_L^2\rangle(\omega)$, is resonantly enhanced due to the resonant pair creation of Josephson polariton pairs and takes the form: 
\begin{equation}
    \langle E_{L,q}E_{L,-q}\rangle(\omega) \rangle = \frac{A}{- \omega^2 + i \gamma \omega + \omega_L^2(q)}\times \frac{1}{-\omega^2 - i \gamma' \omega + \Omega_d^2},
\end{equation}
which shows a resonance when $\Omega_d = 2 \omega_L(q^*)$. 

The experiment measures the intensity of the SHG light and therefore directly $E_L^2$ which is modulated at the frequency $2\omega_L \approx \Omega_d$, i.e. at 3 THz, in agreement with experiment. Note this frequency is independent of temperature, which is also consistent with experiment. 

At the same time as is shown in Fig.~\ref{fig:Parametric}b. the resonant momentum $q^*$ stays approximately the same in the pseudogap state, $T_c < T < T^*$, where $\omega_{p2} = 0$, compared to the superconducting state $T < T_c$ where $\omega_{p2} \neq 0$, as is shown in experiments\cite{vonHoegen22}. This again underlines that the presence of inter-bilayer coherence is not crucial to the phenomenology seen in experiments but intra-bilayer coherence responsible for the non-linear coupling between phonons and Josephson polaritons is.  

As explained in Ref.\citenum{taherian24}, the parametric driving of Josephson polaritons creates coherence in the squeezing part of lower Josephson plasmons, $ \langle a(q) a(-q) \rangle$. In other words, the pump pulse sets the phase of plasmon pairs, whereas individual plasmons, $a(q)$ and $a(-q)$ have random phases, as shown in the schematic of Fig.~\ref{fig:Parametric}d. The experimental signature is able to capture the emitted Josephson polaritons in SHG because the oscillation at $\Omega_d$ is sensitive only to the intensity and not to the phase of the oscillation.


\subsection{Photo-induced c-axis Terahertz Reflectivity}

The appearance of a pair of amplified modes at $\omega_L(\pm q)$ leads to a change in the reflectivity because these modes will couple  with the probe light and show resonant behavior. This effect has been analyzed previously in Ref.\citenum{michael2020parametric} and \citenum{Marios22}. Here, we briefly review the key concepts and outline the theoretical approach. 

The idea is to combine the usual Fresnel solution of the reflectivity problem with the Floquet description of an active medium with an amplified mode. In the conventional Fresnel analysis, one considers the boundary between vacuum and a medium, ensuring that both the electric and magnetic fields remain continuous across the interface. Instead of solving for the eigenmode frequency at a given wavevector $q$, we fix the frequency, $\omega$ of the incoming probe light and solve for a complex wavevector, $q(\omega)$. This approach allows us to determine how incident light at a specific frequency, $\omega$ is transmitted and reflected.

In our case, the medium is described by a model in which Josephson polaritons at frequency $\omega$ are coupled to modes at the idler frequency, $\omega_{id} = \Omega_d - \omega$. We extend the Fresnel approach by imposing boundary conditions at both $\omega$ and $\omega - \Omega_d$. To determine light propagation in the medium we fix $\omega$ and $\omega - \omega_{id}$ and derive complex wavevectors $k_+(\omega,\Omega_d)$ and $k_-(\omega, \Omega_d)$ of the Floquet Josephson polaritons. This allows us to determine the reflectivity at frequency $\omega$ due to an incident probe light at $\omega$ as a function of the driving amplitude. The strength of the coupling $\lambda$ between the modes scales with the square of the drive amplitude, because the relevant interaction given after Eq.~\ref{eq26} term depends on the product of two driven phonon amplitudes, $Q_1$ and $Q_2$. Physically, this means that near parametric resonance, the electric and magnetic fields in the medium become strongly amplified, significantly affecting the reflected probe light.

Now we consider a medium described by Eq. \ref{matricx2by2} where Josephson polaritons at frequency $\omega$ are coupled to $\omega-\Omega_d$. Instead of solving for the Floquet eigenmodes with fixed momentum $q$, we extend the Fresnel procedure to match the boundary conditions for fixed frequencies $\omega$ and $\omega-\Omega_d$ and solve for the momentum $q$ which is now a complex number. In this way the relationship between transmitted, reflected and incident beams at frequency $\omega$ is determined and the reflectivity is calculated for a given drive amplitude. The square of the drive amplitude enters in  the coupling $\lambda$, because the coupling term given after Eq. \ref{eq26} depends on the product of two phonon amplitudes $Q_1Q_2$. 

The conclusion reached in ref.\citenum{michael2020parametric} is that under strong pumping a reflectivity edge that resembles a plasma edge occurs on parametric resonance near $\omega=\Omega_d/2$, i.e. at 1.5 THz, using the parameters used in the present paper. They also showed that the resulting reflectivity features have a universal lineshape, determined solely by the frequency, the effective driving strength, and the effective dissipation in the Floquet medium. In the case of YBCO above its critical temperature, where dissipation of Josephson polariton modes is significant, the appropriate response corresponds to what that work termed “Regime II”, which has the form: 
\begin{equation}
    R_{driven}(\omega) \approx R_{equil}(\omega) + C Re\{\frac{e^{i \theta}}{\omega - \Omega_d/2 + i \gamma } \},
\end{equation}
where $C$ is real and $C e^{i\theta}$ is a complex coupling term that is proportional to the intensity of the drive, $|Q_1 Q_2|^2$ and also depends on the equilibrium refractive index, $n$ as described in Ref.~\citenum{Marios22}. $\gamma$ captures the dissipation of modes at frequency $\Omega_d/2$ and is given by the imaginary part of the refractive index, and the group velocity, $v_g$ : $\gamma(\omega) = \frac{\Im\{ n(\omega)\omega\}}{v_g(\omega)}$. In this reference~\citenum{Marios22}, it was shown that the only relevant scale to get the edge like feature reported in experiments is the ratio between $\gamma$ and the driving strength. A sketch of this response that closely resembles experimental data of Ref.~\citenum{vonHoegen22} is plotted in Fig.~\ref{fig:Parametric}c. Note that the Floquet framework leads to a change in the reflectivity which we compare with experiment. While one can uniquely convert this reflectivity to the real and imaginary parts of an effective conductivity $\sigma_{eff}$, it is important to keep in mind that  $\sigma_{eff}$ should not be interpreted as some modified equilibrium conductivity. For example, it does not have to satisfy Kramers-Kronig and its real part can be negative, indicating amplification. As a function of temperature, the effect is reduced in amplitude both because the pairing correlations are reduced therefore diminishing $C$ and because dissipation increases, enhancing $\gamma$.

\section{Conclusions}

This paper examines the dynamics of Josephson plasmon polaritons above $T_c$ and the parametric amplification of the lower plasmon branch under intense driving. While initial observations of reflectivity changes were interpreted from the perspective of equilibrium systems as light-induced plasmon edges \cite{Hu14,Kaiser14}, subsequent research has emphasized the inherently Floquet nature of these photoexcited states \cite{michael2020parametric,taherian24}. Our theoretical framework, while using a slightly modified approach, aligns with the parametric amplification mechanism previously established \cite{michael2020parametric}; however, we depart from earlier works in our interpretation of the results. Whereas earlier models \cite{michael2020parametric,taherian24} suggest that the parametric excitation of lower plasmons implies the existence of in-plane superconductive coherence—either pre-existing or drive-induced—the central thesis of this work is that such an inference is not strictly necessary.
We emphasize the fact that in the case when the Josephson coupling between bilayers is weak, the coupling between bilayer is capacitively driven and the nature of the mode does not depend much on the lower plasmon frequency $\omega_{p2}$ being finite, as long as its energy is not too close to $\omega_{p2}$ (see Fig. \ref{fig:Parametric}). We demonstrate this by setting $\omega_{p2}=0$ and show that the parametric amplification proceeds as before. In this model there is clearly no mechanism for inter-bilayer phase coherence. There is no physical current that flows between bilayers.  What is coherent under the drive is the Maxwell displacement current and that is sufficient for the parametric amplification model to apply.

In addition to the SHG and reflectivity measurements, there was a recent report of the generation of a small additional magnetic field  outside the pumped area that is a small fraction $ \approx 10^{-3}$ of an applied magnetic field and along the same direction\cite{Sebastian24}. This was interpreted as flux exclusion due to creation of a superconducting-like state under pumping. In ref.~\citenum{michael2024giant}
we proposed an alternative explanation due to an instability of the phase variable $\theta_j$ under drive and in the presence of an applied magnetic field. 
The basic assumption of local pairing is the same, but in this case we focus on the direct coupling of the drive electric field to the relative phase between members of a bilayer, rather than through the phonon coupling.
We show that this problem can be described by an effective quasi-static Floquet theory where the Josephson potential becomes effectively inverted leading to a paramagnetic instability. This instability leads to a paramagnetic pulse emitted towards the detector which agrees qualitatively and quantitatively with the observed magnitude. Since this effect happens in parallel with the phonon drive, the reader may be concerned that the inversion of the Josephson potential may invalid the consideration in this paper, which is based on expansion about a stable Josephson effect. In appendix C we show that this does not happen. Even in the presence of a strong electric field that inverts the Josephson potential, a stable low plasmon-polariton mode survives so that the discussions in this paper remains valid.

We should point out that by using ultra narrow pulses that  target individual phonons,  very interesting and rich response in the reflectivity has been reported ~\cite{Liu_20} . However, nonlinear spectroscopy such as SHG has not been reported with these narrow pulses. The set-up in this paper cannot address these data. On the other hand, application of our framework to the original 3 way mixing model of ref.~\citenum{vonHoegen22, michael2020parametric} where a single phonon mode couples to an upper and lower plasmon may be useful here. 

There is also a recent report of in-plane response under a out of plane drive.~\cite{rosenberg25}. This problem will be addressed in a separate paper. 

We conclude that the local pairing  model successfully captures properties of the photoexcited state, including THz reflectivity, time-resolved second harmonic generation, and the dynamics under an external magnetic field. Since the bilayer structure of YBCO plays a crucial role, our theory predicts that similar phenomena should not occur in monolayer systems such as Hg2201 and LSCO. It will be interesting to test this in future experiments.

When considering earlier experimental results, we point out that the phase of an individual layer may have very short correlation length $\xi$ (say of the order of a few nanometers), so that its effect on the overall transport properties is challenging to detect with conventional probes. 
Here we comment on the consistency of our picture with measurements on physical properties that are sensitive to the average phase difference $\theta_{s,j}$ of bilayer $j$ which generally find that superconducting fluctuation effects diminish to varying degrees above $T_c$ and become unobservable much below $T^*$.  A number of experiments such as microwave conductivity \cite{corson1999vanishing}, have shown that phase coherence does not extend to more than 1.2$T_c$. However, other experiments such as Nernst effect and diamagnetism\cite{li2010diamagnetism} find detectable fluctuation effects up to about 2$T_c$ in underdoped cuprates. A similar conclusion is reached for fluctuation conductivity including the presence of a magnetic field\cite{rullier2011high}. However, these temperature scales are much below $T^*$. Another interesting experiment is by Bergeal et al.\cite{bergeal2008pairing} who used the underdoped YBCO as one electrode of a Josephson junction to probe the time scale of the pair  susceptibility. They found a sharp peak which diminishes above $1.2 T_c$, which supports a rapid suppression of pair fluctuations. On the other hand, in a follow-up experiment, Koren and Lee\cite{koren2016observation} found that this sharp peak sits on top of a broader peak that persists up $90K$ which is the limit of the experiment set by the transition temperature of the opposing electrode. This can be taken as evidence for pair fluctuations up to at least this scale. In this connection, it is interesting to note that measurements on the c-axis conductivity have shown that the superfluid density has a long tail that extends up to $160K$, even though its value has decreased to a few percent ~\cite{TajimaPRL}. Since the c-axis superfluidity requires phase coherence between bilayers, this indicate the existence of some inter-bilayer coherence of the average phase $\theta_{s,j}$ up to $160K$ in underdoped YBCO. On the other hand, as mentioned earlier, there is strong evidence for phase coherence between members of the bilayer based on  the observation of transverse Josephson plasmon up to $180K$,\cite{dubroka2011evidence}. This observation has been confirmed by an independent method of monitoring the temperature dependence of a phonon mode\cite{TajimaPRL}. Thus there are indeed strong indications of local pairing persisting to far above $T_c$, but these measurements may not have the sensitivity to detect very short correlation length that may exist above $180K$. In this sense, subjecting the samples to strong pump may be the new tool that can open a new window to this possibility.

In our picture the relative phase between members of a bilayer is assumed to have much longer correlation length than $\xi$ and persists up to $T^*$. For the paramagnetic magnetic field instability we require the correlation length to be longer than the magnetic field screening length $\lambda_s$ at the edges of an isolated bilayer, which can be quite long, on a scale of several hundreds of nanometers. It is quite possible that the parametric amplification is more forgiving in that different patches of the samples with phase difference of multiple of $2\pi$ may be amplified coherently. In any event, the final message is that the minimal requirement to explain all the existing data is for the existence of a local pair phase at equilibrium up to the pseudogap temperature $T^*$. 

The quest for understanding the pseudogap in cuprate superconductors has a long history. It is considered to be a crucial part of the high-$T_c$ puzzle. 
An important question which remains unsettled is the origin of the energy gap itself.
Much of the discussions has centered on anti-ferromagnetic spin fluctuations, identifying the gap as a spin gap due  to 
the suppression of spectral weight due to strong correlation near the Mott transition\cite{wu2018pseudogap}. Other scenarios, such as orbital currents  \cite{varma2019pseudogap} or  pair density waves \cite{lee2014amperean,agterberg2020physics,zheng2025competing} have also been proposed. Another set of ideas comes from the resonating valence bond (RVB) scenario 
, where   the pairing of spinons lead to the opening of a gap in the spin excitation spectrum in mean field theory\cite{kotliar1988superexchange, anderson2004physics}. This point of view has been extended beyond mean-field, and it was argued that there is a regime where the spinon pairing inherits properties of electron pairing, resulting in a strongly fluctuation pairing state that 
extends up to $T^* > T_c$  ~\cite{senthil2009coherence}. Note that the energy gap associated with the pseudo-gap is generally very large, of the order of $100 meV$, and a number of experiments point out that it is not simply connected to the superconducting gap, setting up the one gap vs two gap debate.\cite{le2006two}. In this paper we do not take a stand on this debate. To satisfy the assumption of local pairing amplitude, the pairing can be a fluctuating pair state different from the d-wave pair, or the pairing may co-exist as a fluctuating part of some more complicated "interwined" order.  Indeed, recent gauge theory based descriptions point to this direction\cite{pandey2025thermal}. Along the same vein, the pseudo-gap temperature $T^*$ is often considered to be a true phase transition, with onset of  properties such as nematicity \cite{sato2017thermodynamic} or orbital loop currents \cite{bourges2021loop}. Our model requires only a cross-over at a similar energy scale. The relationship of such cross over to a possible phase transition is beyond the scope of our phenomenological model.  With these disclaimers, our conclusion remains that the notion that 
local pair amplitude exists up $T^*$ has a strong impact on the understanding of the pseudogap. From this point of view, the series of ultra fast pumping probe experiment may have unveiled an important property of the pseudogap state that has escaped detection by traditional linear response type of techniques. This being said, we re-iterate that our work neither rules out or rule in the scenario where coherent superconductivity is created by pumping. The goal of this paper is to offer an altenative explanation of the data.

Finally, we note that a recent low-temperature local shot noise measurement using scanning tunneling microscopy (STM) has inferred that the large pseudogap is a pairing gap \cite{niu2024equivalence}. The large size of the observed gap (30 to 70 meV) suggests that it should survive up to a high temperature, which aligns with the picture presented in this paper. 

\paragraph{Acknowledgements:} We acknowledge multiple fruitful discussions with Andrea Cavalleri and members of his group, collaboration on a related subject with Duilio de Santis, and stimulating discussions with Gil Refael.  M.H.M. is grateful for the financial support received from the Alex von Humboldt postdoctoral fellowship and the hospitality of the Max Planck Institute for the Physics of Complex systems. E.A.D. acknowledges support from ETHZ, the SNSF project 200021\textunderscore212899, the ARO grant number W911NF-21-1-0184, and the SNSF Sinergia grant CRSII--222792. P.A.L. acknowledges support from DOE (USA) office of Basic Sciences Grant No. DE-FG02-03ER46076.

\section{Supporting Information}

\appendix
\section{ Appendix A : The Josephson plasmon polariton modes.}

The Josephson plasmons are admixed via the Maxwell equation to the propagation of light, and is more properly called the Josephson plasmon polariton (JPP) modes.
In this section we give a detailed derivation of JPP mode for a structure consisting of stacked bilayers that is appropriate for YBCO. We determine the mode structure by solving the Maxwell equations with the assumption that the in-plane currents and charge densities are confined within the cuprate layers while Josephson currents flow between the layers. It is worth noting the difference from the earlier treatment given by Michael et al~\cite{michael2020parametric}. There the Maxwell equation was formulated in terms of the vector potential $\textbf{A}$ and the approximation was made that the in plane component of $\textbf{A}$ is confined to the layers. Here we do not make this assumption. It turns out that the formulation in terms of $\textbf{E}$ is more amenable to analytic treatment and we can solve the full three dimensional structure of $\textbf{E}$ for the case when the momentum perpendicular to the plane $q_z$ is zero.

\begin{figure}[t!]
    \centering
    \includegraphics{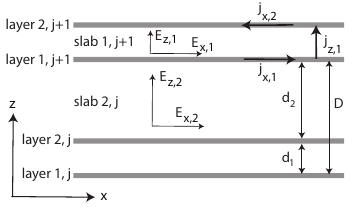}
    \caption{The labeling scheme for the bilayer structure. Each bilayer is labeled by the index $j$ with the bottom and top layer labeled by 1 and 2. The in-plane super-currents tend to flow in opposite directions.Our model assumes  Josephson current between members of the bilayer but none between bilayers. Neverthelss, electric fields are present between bilayers in slab 2 due to capacitive coupling.}
    \label{fig:Bilayer}
\end{figure}

The in-plane current and density is given by:
\begin{subequations}
\begin{align}
    \vec{j}(\vec{x},z,t) = \sum_{\mathclap{\substack{\alpha = \mbox{layer},\\ j = \mbox{unit cell}}}} \delta(z - z_{\alpha,j}) \vec{j}_{\alpha,j} (\vec{x},t),\\
    \rho(\vec{x},z,t) = \sum_{\mathclap{\substack{\alpha = \mbox{layer},\\ j = \mbox{unit cell}}}} \delta(z - z_{\alpha,j}) \rho_{\alpha,j} (\vec{x},t),
\end{align}
\end{subequations}
where $\alpha=1,2$ is the layer index and $z_{\alpha.j}$ the position of layer $\alpha$ in init cell $j$ along the z-axis. 
As sketched in Fig.~\ref{fig:Bilayer}, in YBCO the structure consists of bilayers with spacing $d_1$ which are separated from each other by a larger spacing $d_2$. The unit cell dimension is $D=d_1+d_2$. We shall use the layer index $\alpha=1,2$ to denote the bottom and top members of a bilayer. We shall refer to the space between members of a bilayer as slab 1 and the space between the bilayer as slab 2. 
The Josephson current along the z-axis runs between two layers: 
\begin{align}
\begin{split}
    j_{z,1,j} (\vec{x},t) =& e^* j_{c,1} \sin \Bigg( \phi_{2,j} - \phi_{1,j} \\
    &- e^* \int_{z = z_{1,j}}^{z = z_{2,j}}dzA_z(\vec{x},z,t )\Bigg),
    \end{split}\label{current1}
    \\ 
\begin{split}
    j_{z,2,j} (\vec{x},t) =& e^* j_{c,2} \sin \Bigg( \phi_{1,j+1} - \phi_{2,j} \\
    &- e^* \int_{z = z_{2,j}}^{z = z_{j,lj+1}}dzA_z(\vec{x},z,t )\Bigg),
\end{split}\label{current2}
\end{align}
where $j_{c,\alpha}$ denotes the Josephson current across slab $\alpha$. There can be an additional normal current component due to normal carriers, especially for $T>T^*$ as well as due to motion of Josephson vortices. These will add a damping term to the equation of motion below, but these will be left out for simplicity. The addition of a damping term will not significantly modify the main conclusion concerning parametric amplification reached below except for allowing the system to reach a stead state.

Combining  the Maxwell equations for $\nabla \times \mathbf{E}$ and $\nabla \times \mathbf{B}$ we eliminate the magnetic field to obtain the equation for the electric field  :
\begin{align}
    \frac{1}{c^2} \partial_t^2 \textbf{E}(\vec{x},z,t) = \nabla \times \nabla \times \textbf{E} - \mu_0 \partial_t \textbf{j}(\vec{x},z,t)
\end{align}
Writing out the components along $z$ and $x$ :
\begin{align} 
   \begin{split}\frac{1}{c^2} \partial_t^2 E_z(\vec{x},z,t) - \partial_x^2  E_z(\vec{x},z,t) + \partial_z  \partial_x E_x(\vec{x},z,t) = & \\
   - \mu_0 \partial_t j_z(\vec{x},z,t)&,\end{split}\label{Maxwell1} \\
 \begin{split}\frac{1}{c^2} \partial_t^2 E_x(\vec{x},z,t) - \partial_z^2  E_x(\vec{x},z,t) + \partial_x  \partial_z E_z(\vec{x},z,t) =& \\ - \mu_0 \partial_t j_x(\vec{x},z,t)&,
\end{split}
\end{align}
Linearizing Eq.\ref{current1} , we obtain the standard AC Josephson relation
\begin{align}
    \partial_t j_{z,1,j} (\vec{x},t) =&(e^*)^2 j_{c,1} \int_{z = z_{1,j}}^{z = z_{2,j}}dzE_z(\vec{x},z,t ), \label{jc1}\\
    \partial_tj_{z,2,j} (\vec{x},t) =& (e^*)^2 j_{c,2}   \int_{z = z_{2,j}}^{z = z_{1,j+1}}dzE_z(\vec{x},z,t ),
     \label{jc2}
\end{align}
The integral is simply the voltage drop across the slab. For the in-plane current, we take the phenomenological form after Fourier transform to frequency domain:
\begin{align}
  \vec{j}_{i,j} (\vec{x},\omega) = \sigma(\omega)  \vec{E}_{i,j} (\vec{x},\omega).  \label{jplane}
\end{align}
where $\sigma$ is the two dimensional conductivity which we take to be the Drude form,
\begin{align} 
  \sigma(\omega) = \frac{e^2 n_{2D}/m}{-i\omega + \Gamma}.
  \label{drude}
\end{align}
If the damping rate $\Gamma =0$ we have a superconducting response in the layer, but we will keep $\Gamma$ finite because the layers are metallic in the pseudogap phase. Alternatively we can also consider a two fluid model where $\sigma$ is a sum of a superfluid term $n_s/im\omega$ and a normal fluid term of the form given by Eq. \ref{drude}. Below we will proceed with a general $\sigma$. We will find that its form has minimal effect on the JPP dispersion in YBCO.

We supplement Eq ~\ref{Maxwell1} with Coulomb law which gives the discontinuity of electric field $\Delta E_{z,j}$ across layer $j$ in terms of the charge density. Together with the continuity equation, we find for layer 1
\begin{align}
    \partial_t \Delta  E_{z,1} =  ((j_2-j_1)-\partial_x j_{1,x})/\epsilon_0. \label{discon}
\end{align}
and similarly for layer 2. Equations ~\ref{Maxwell1} to \ref{discon} fully determine the propagation of the electromagnetic wave through the sample.

Next we Fourier transform the equations along the planar  direction and label the Fourier component by $q$ which we take to be along the $x$ direction. The Maxwell equation takes the simpler form:
\begin{align} \label{max1}   
(\omega^2-c^2q^2)E_z=iqc^2\partial_z E_x -i\omega \mu_0 j_z(q,\omega)
\end{align}
\begin{align} \label{max2}
    \omega^2E_x + \partial_z^2E_x=iqc^2\partial_z E_z -i\omega \mu_0 j_x(q,\omega)
\end{align}
We will consider waves which propagate along the plane, i.e. with $q_z=0$, in which case $E_x$ ,$ E_z$ and the currents are periodic in the $z$ direction, with the period of the unit cell. We denote the electric field in slab $i$ by $E_{z,i}$ and $E_{x,i}$ which depends on $z,q, \omega$ and the currents by $j_{z,i}$ across slab $i$ and $j_{x,j}$ in layer $j$. The currents depend only on $q, \omega$ and are independent of $z$. The full equations are solved analytically in Appendix B. Here we describe an approximate solution which assume that $E_{z,i}$ is constant in the z direction and the term $\partial_x j_{1,x}$ in Eq. \ref{discon} can be neglected for small $q$ . This recovers the standard dispersion for the Josephson plasmon polariton in the literature and is sufficient for the purpose of this paper. This is because we show in  Appendix B that the neglected terms are controlled by the parameter $\beta = \frac{\left(e^*\right)^2 n_s d_1 }{2 \epsilon_0 m c^2}$. Interestingly the same parameter enters the fluctuation 
contribution to the diamagnetic susceptibility as $\chi_s=\beta/3$  \cite{michael2024giant} and $\chi_s$ is estimated to be $\approx 10^{-5}$. Therefore this approximation is fully justified for YBCO. 

We denote  $E_{z,i}$ by a constant $E_i$ in slab $i$. To simplify notation, we drop the subscript $z$ from the electric field and current along $z$ from now on. By integrating Eq. \ref{max1} over the unit cell along $z$ we find:
\begin{align}  \label{E1}
    (\omega^2-c^2q^2)(d_1E_1+d_2E_2)=-i\omega\mu_0((d_1j_1+d_2j_2)
\end{align}
Importantly the term involving $\partial_z E_x$ vanishes due to the periodic boundary condition. Therefore within this approximation, the behavior of the current along $x$ does not matter, whether it is dissipative or superfluid like. Furthermore, Eq.\ref{jc1} becomes 
\begin{align} \label{j1E}
    j_1=\omega_{p1}^2 E_1/(-i\omega)
\end{align}
where the Josephson plasmon frequency $\omega_{p1}$ for slab $1$ is given by
\begin{align}
    \omega_{p1}^2= e^* j_{c,1} d_1.
\end{align}
and similarly for slab 2. Eq. \ref{E1} becomes
\begin{align}  \label{E2}
    (\omega^2-c^2q^2)(d_1E_1+d_2E_2)= \omega_{p1}^2 d_1 E_1 + \omega_{p2}^2 d_2 E_2
\end{align}
This equation is supplemented by the equation for the discontinuity of the electric field across the layer. As stated earlier, the approximation involves ignoring the term $\partial_x j_{1,x}$ in Eq.\ref{discon}. Combining with Eq.\ref{j1E} we find
\begin{align}   \label{discon2}
    \omega^2(E_1-E_2)=\omega_{p1}^2E_1 - \omega_{p2}^2E_2
\end{align}
This equation fixes the ratio between $E_1$ and $E_2$. Plugging into Eq. \ref{E2}, we find the dispersions for the upper and lower JPP modes $\omega_U$ and $\omega_L$ are given by
\begin{widetext}
\begin{align}
\label{dispersion}
    \omega_{U,L}^2= (\omega_{p1}^2+\omega_{p2}^2+c^2q^2)/2 \pm 1/2 \sqrt{c^4q^4+2c^2q^2(\omega_{p1}^2+\omega_{p2}^2-2\omega_{pT}^2)+(\omega_{p1}^2-\omega_{p2}^2)^2} 
\end{align}
\end{widetext}
where the transverse plasmon frequency is given by
\begin{align}
\label{transverse}
   \omega_{pT}^2=(d_1/D)\omega_{p2}^2+(d_2/D)\omega_{p1}^2
\end{align}

A sketch of the JPP dispersion is shown in fig. \ref{fig:plasmon}b.



\appendix
\label{Appendix B}
\section{ Appendix B : Further refinement of the Josephson plasmon polariton modes: inclusion of x component of electric field.}

\begin{figure}
    \centering
    \includegraphics[width=\linewidth]
{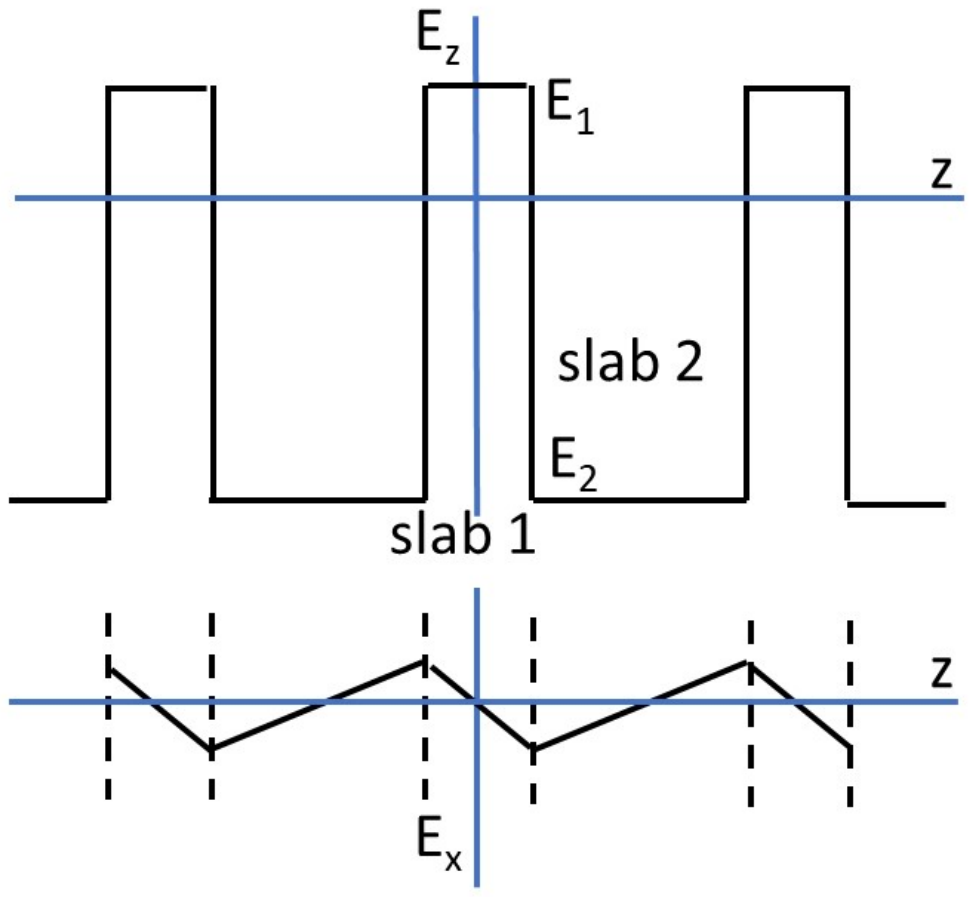}
\caption{ A sketch of the electric field $E_z$ (top) and $E_x$(bottom) across different layers along the $z$ direction. $E_z$ is approximately constant in each slab with a discontinuity across each layer, while $E_x$ is approximately linear inside each slab with a slope discontinuity across each layer. }\label{Ex}
\end{figure}

In this Appendix we relax the approximations made in the main text. We compute the $x$ component of the electric field which we label by $E_{x,j}$ for slab j. Note that unlike the $z$ component, the $x$ component is continuous across the layer, but its derivative along $z$ may be discontinuous. See Fig. \ref{Ex}. Within  slab 1, Eq. \ref{max1} and \ref{max2} become
\begin{align} \label{max3}
    (\omega^2-c^2 q^2) E_{z1} = iq c^2 \partial_z E_{x1} - \omega_{p1}^2 E_{z1}
\end{align}
\begin{align} \label{max4}
    \omega^2 E_{x1}-c^2 \partial_z^2 E_{x1} = iqc^2\partial_z E_{z1}
\end{align}
and similarly for slab 2. Note that there is no current contribution to the RHS because the current along $x$ is restricted to the layers. Solving Eq. \ref{max3} for $E_{z1}$ and plugging into Eq. \ref{max4} we obtain an equation for $E_{x1}$:
\begin{align}
    c^2\partial_z^2 E_{x1}= -\kappa ^2 E_{x1}
\end{align}
where
\begin{align}
    \kappa^2=\frac{\omega^2/c^2}{1+\frac{q^2c^2}{\omega^2 - (\omega_{p1}^2+c^2q^2)}}
\end{align}
Using the symmetry about the midpoint of each slab and measuring $z$ from it, we find the solution 
\begin{align}
    E_{x1}=(s_1/|\kappa|) \sin(\kappa z)\\
    E_{x2}=(s_2/|\kappa)|| \sin(\kappa z)
\end{align}

Note that $\kappa^2$ is positive for $\omega < \omega_{p1}$ and negative for $\omega > \omega_{p1}$. On the other hand, $|\kappa|$ is of order $q$ so that $|\kappa| d_2 << 1$ and we can simply write $E_x$ as a linear function with slops $s_1 $ and $s_2 $. Being periodic over the unit cell, we have $s_1d_1+s_2d_2=0$, so that $s_1/s_2=-d_2/d_1$ . It remains to determine the individual slope $s_1$. This is set by the boundary condition across a layer. At layer 1, Eq. \ref{max4} takes the form:
\begin{align}
    \omega^2 E_{x1}+c^2 \partial_z^2 E_{x1} = iqc^2\partial_z E_{z1} -   i \delta(z-z_0)\omega \mu_0 \sigma(\omega) E_{x1}
\end{align}
Note that we restored the in-plane current as a delta function in the layer at position $z_0$. Recall that $E_{z1}$ has a discontinuity $\Delta E_{z1}$ so that the first term on the RHS is also a delta function. In comparison, the first term on the LHS can be dropped. The second term shows that the $z$ derivative of $\partial_z E_{x1}$ is proportional to a delta function, whose coefficient gives the discontinuity   in $\partial_z E_{x1}$ which is $s_1-s_2$. We find,
\begin{align}
    c^2(s_1-s_2)=iqc^2\Delta E_{z1}-i\omega \mu_0 \sigma(\omega) (-d_1/2)s_1 
\end{align}
Using the condition  $s_1/s_2=-d_2/d_1$ derived earlier, we find
\begin{align} \label{s1}
    s_1= \frac{iq\Delta E_{z1}}{(1+d_1/d_2)-\beta}
\end{align}
where
\begin{align}
    \beta=-i\omega\mu_0\sigma(\omega)d_1/2
\end{align}

This gives the solution for the in-plane electric field $E_{x1}$ and $E_{x2}$. Now we go back to check the validity of the assumption that the $z$ components $E_{zi}$ is independent of $z$. Going back to Eq. \ref{max1}, we see that the spatial dependence comes from the term $iqc^2\partial_z E_x$ which is proportional to $\cos(\kappa x)$. Since $\kappa d_2 \ll 1$ this correction is independent of $z$. Hence the assumption of constant $E_z$ remains valid and we can denote them by $E_1$ and $E_2$ in slab 1 and 2 as before. Thus Eq. \ref{E2} remains valid and the mode frequencies are solved by combining this with the modification of Eq. \ref{discon2}.  Eq. \ref{discon} gives the discontinuity of $E_z$ across layer 1:
\begin{align}
    -i\omega \Delta E_{z1}= (\omega_{p1}^2E_1-\omega_{p2}^2E_2)/(i\omega) - iq\mu_0\sigma E_{x1}
\end{align}
where in the last term on the RHS $E_{x1}$ is evaluated at layer 1 and takes the value $-s_1d_1/2$. Using Eq. \ref{s1} we find
\begin{align}  \label{discon3}
    \omega^2 (E_1-E_2)= \frac{(\omega_{p1}^2E_1-\omega_{p2}^2E_2)}{1-\frac{\beta q^2c^2/\omega^2}{1+d_1/d_2-\beta}}
\end{align}
This equation gives the correction to Eq. \ref{discon2} and fully account for the effect of $E_x$. The JPP mode frequency is obtained by combining Eqs. \ref{E2} and \ref{discon3}. Since $q^2 c^2 \approx \omega^2$ we clearly see that the dimensionless parameter $\beta$ serves as the control parameter of this correction. When $\beta$ is small the correction to the standard dispersion is negligible.


We now provide a rough estimate of $\beta$. 
Substituting  $\sigma(\omega)=e^{*2}n_s/i\omega m$ into  $\beta$,
where $n_s$ is a 2D local superfluid density.  we find that the same $\beta$ appears in the diamagnetic susceptibility of an isolated bilayer.  This is found to be very small for YBCO,  ~\cite{michael2024giant} ($\beta \approx 3.10^{-5}$)  as mentioned in the text.

For completeness it is interesting to consider the limit $\beta \gg 1$. In that case it is easy to show that the mode frequencies become $\sqrt{\omega_{p1}^2 + q^2c^2}$ and $\sqrt{\omega_{p2}^2 + q^2c^2}$. These are just the decoupled Josephson plasmons propagating in the two slabs. In this limit the conductivity within the layers dominates and screens the electromagnetic field, decoupling the layers. In this paper we assume that $\beta \ll  1$ so that the modification to the standard JPP dispersion is very small, as supported by experimental evidence.

\section{Appendix C: Lower plasmon in the presence of an unstable upper plasmon}

\begin{figure}
    \centering
    \includegraphics[width=\linewidth]{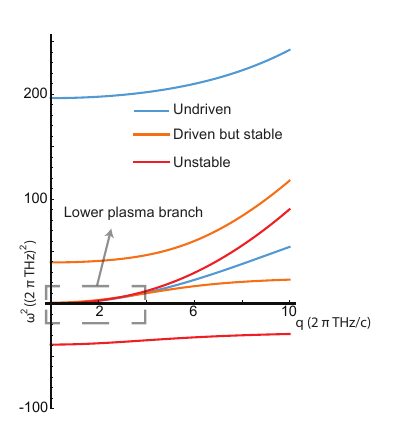}
\caption{\textbf{Plasmon dispersion in the presence of unstable upper plasmon:} We plot the dispersion relation in the pseudogap phase of YBCO for the undriven case : $\omega^2_{p1,eff} = \omega^2_{p1}$} (blue), driven but not unstable, $0 < \omega_{p1,eff}^2 = 0.2 \omega_{p1}^2 < \omega_{p1}^2$ (orange) and the unstable case $\omega_{p1,eff}^2 = - 0.2 \omega_{eff}^2$ (red). In all three cases we find well defined lower plasmon dispersions that are largely unaffected by the instability of the upper plasmon. 
\label{fig:drivenplasmon}
\end{figure}

In this Appendix we demonstrate how the mechanism for giant dynamical paramagnetism presented in Ref.~\citenum{michael2024giant}, is consistent with the current work and as a result the two papers provide a consistent model that can account for most of the available data. In that reference we argued that large electric field driving can "flip" the Josephson junction potential between members of the bilayer : 
\begin{equation}
    \omega_{p1,eff}^2 \approx \omega_{p1}^2\int_0^{\frac{2 \pi}{\omega_d}} dt \left( \cos( A \cos(\omega_d t)   \right) = \omega_{p1}^2 J_0(A).
\end{equation}
where $A$ is the amplitude of the homogeneous oscillations of the excited phase difference between the two layers and $\omega_d \sim 17 - 20 $ THz, which is faster than the Josephson plasma frequencies and therefore we can consider the effective time average effect of this drive (for a more complete picture consult Ref.~\citenum{michael2024giant}). We argued in that reference that for large enough $A \sim \pi $, the effective Josephson coupling flips sign and $\omega_{p1,eff}^2 < 0 $ , which leads to an instability. We argue that this instability is the cause of the magnetic field response observed in recent experiments in pumped YBCO\cite{Sebastian24}. Here we show, that even at very large amplitudes where this instability is present we can still define lower plasmons, that have almost the same dispersion as before. As a result, we would expect that parametric resonant phenomena discussed in this article can co-exist with this intra-bilayer instability explain the magnetic field results. From Eq.~(\ref{eq25}) and (\ref{eq26}), the dispersion relation of the two plasmonic branches is given by the secular equation:
\begin{equation}
    \begin{pmatrix} - \omega^2 + \omega_{p1,eff}^2 +v_1^2 q^2 && v_1^2 q^2 \\
    v_1^2 q^2  && - \omega^2 + v_2^2 q^2 \end{pmatrix} \cdot \begin{pmatrix} E_{1}(q,\omega) \\E_2 (q,\omega) \end{pmatrix} = 0 .
\end{equation}
We find that there is always a gapless solution for $q = 0$, that we can assign to the lower plasmon, $\omega_L(q = 0) = 0$ and a gapped solution that undergoes the instability as $\omega_{p1,eff}^2$ changes sign, with frequency $\omega_{U}^2 =\omega_{p1,eff}^2$. In Fig.~\ref{fig:drivenplasmon}, we show that the lower plasmon dispersion is largely unaffected by the instability of the upper plasmon. As a result while the instability might cause some quantitative changes to the parametric process, we expect resonant frequencies and momenta from the phonon-lower plasmon dynamics to not be significantly affected.

\bibliography{references}{}
\end{document}